\documentstyle[12pt]{article}
\begin{document}
\begin{titlepage}
\vspace{2.5cm}
\begin{centering}

{\LARGE Self-Forces on Electric and Magnetic}\\
{\LARGE Linear Sources in the Space-Time}\\
{\LARGE of a Cosmic String}\\
\vspace{1cm}
{\large E. R. Bezerra de Mello$^{\ast}$ and V. B. Bezerra$^{\dagger}$}\\
{\large Departamento de F\'{\i}sica, UFPB, 58051-970, J.Pessoa,PB,}\\
{\large BRAZIL}\\
{\large C. Furtado and F. Moraes$^{\diamond}$}\\
{\large Departamento de F\'{\i}sica, UFPE, 50670-901, Recife, PE,}\\
{\large BRAZIL}

\end{centering}
\vspace{1.5cm}

\begin{abstract}

In this paper we calculate the magnetic and
electric self-forces, induced by the conical
structure of a cosmic string space-time,
on a long straight wire which presents either a constant
current or a linear charge density.  We also show how these self-forces are
related by a Lorentz tranformation and, in this way, explain what two different
inertial observers detect in their respective frames.

\end{abstract}
\hspace{0.8 cm}
PACS numbers: 98.80.cq, 04.40.+c, 41.20.-q

\vspace{2.0cm}

\noindent
$^{\ast}$e-mail: cendfi50$@$brufpb.bitnet\\
\\
$^{\dagger}$Present address:\hspace{3.5cm}$^{\diamond}$Present address:\\
Blackett Laboratory\hspace{3cm}School of Natural Sciences\\
Imperial College\hspace{3.7cm}Institute for Advanced Study\\
London SW7 2BZ\hspace{3.4cm}Princeton, NJ 08540\\
UK\hspace{6cm}USA

\end{titlepage}

\baselineskip = 18pt

 It is well known that the space-time produced by a
thin, infinite, straight cosmic string, has no Newtonian
potential [1] and cannot induce curvature
(locally the curvature vanishes everywhere except at the
source). However, there are some
global non-trivial topological effects associated with
this space-time which
can be measured [2].

Recently, Linet [3] and Smith [4] have
shown that a charged point particle placed in a static
space-time, produced by an infinite straight cosmic string,
becomes subjected to a finite repulsive electric self-force.
This phenomenon is a consenquence of the distortion in the
particle field caused by the lack of global flatness
of this space-time.

The purpose of this paper is to investigate how the self-force
phenomenon manisfests in the magnetic case; i.e.,
we  study what happens either to an electric current
or to a linear charge distribution in the conical space-time produced by a
cosmic string.
Furthermore, we show that the respective
self-forces are related by a Lorentz transformation (a
boost along the  direction of the string). In
order to analyse this problem we shall adopt, for
simplicity only, the case where the magnetic and electric  sources are both
infinite straight wires parallel
to the cosmic string, which we assume lies along the
z-axis. Before we start our calculation, we add
that the analysis presented here can also be applied
to some elastic solids which present a specific kind
of linear defects named disclinations [5].
As shown by Katanaev and Volovich [6],
the change in the topology of the elastic medium produced  by this linear
defect is the same as the one produced by a cosmic string in Minkowski
space-time.
This similarity goes beyond topology: for some
applications both kinds of defects can be treated
using the same geometrical approach; i.e., their surrounding
space-time  can be described by similar geometrical
points of view. In this approach, the line element in the section
z=t=constant, is given by

\begin{equation}
d\vec\ell^2_{(2)} = dr^2 + r^2d\theta^2 ,
\label{1}
\end{equation}
where $r\geq 0$ and $0\leq\theta\leq 2\pi\alpha$,
being $\alpha=1/p$ a parameter related to the
linear mass density $\mu$ of the cosmic string
by $\alpha=1-4\mu G/c^2$. For an elastic solid
$\alpha=1+\lambda/2\pi$,
where $\lambda$ is the angle which defines the wedge [7]
For the cosmic string, $\alpha$ is smaller than 1.
For disclinations, however, it can also assume
values greater than 1, which correspond to
an anti-conical space-time with negative curvature.

Before going on into the calculations let us say a few words concerning the
motivation to study this problem. Certainly, the understanding of the
interaction between currents and a topological defect, such as a cosmic string,
can give us some informations on the behavior of a superconducting string [8]
in the space-time of the defect. Also, at a low energy level, it may be useful
in the study of electromagnetic properties of disclinated media [5,9].

Now, after this brief introduction about our problem,
let us first analyse the easiest case of a linear charge distribution parallel
to the string.

The electrostatic potential energy for an arbitrary
charge distribution is:

\begin{equation}
U_{Ele}={1/2}\int{ d^3 r\rho(\vec r)\Phi(\vec r)},
\label{2}
\end{equation}
where $\rho(\vec r)$ is the charge density and $\Phi(\vec r)$
the scalar potential. Because we are interested in
studying the problem of the electrostatic self-energy for
a constant linear charge density parallel to the string, the Maxwell equation
for the scalar potential

\begin{equation}
\nabla^2\Phi(\vec r)=-4\pi\rho(\vec r),
\label{3}
\end{equation}
reduces to a two-dimensional one.
In this case,
we can write the  electrostatic potential energy per unit length as

\begin{equation}
{U_{Ele}\over \ell}={1/2}\int\int d^2r d^2r'\rho(\vec r)G_p^{(2)}(\vec r,
\vec r \,')\rho(\vec r \,'),
\label{4}
\end{equation}
where $G_p^{(2)}(\vec r,\vec r \,')$ is the two-dimensional
Green's function on the conical two-geometry.

The analogue magnetic expression for the linear energy density can be writen
for the particular configuration where the
constant current is also paralell to the topological
defect. For this situation we have

\begin{equation}
{U_{Mag}\over \ell}=\frac{1}{2c^2}\int\int d^2r d^2r'\bigl({\vec J}(\vec r)
\cdot{\vec J}(\vec r \,')\bigr)G_p^{(2)}(\vec r,\vec r \,').
\label{5}
\end{equation}
(In the derivation of (5), we  assumed that the
vector potential $\vec A$ associated with the magnetic
field $\vec B$ is in the Coulomb gauge; i. e.,
$\vec\nabla\cdot \vec A=0$).

Now, by using equations (4) and (5), we  obtain the linear
self-energy densities, admiting for the electric and
magnetic sources the expressions below:

\begin{equation}
\rho(\vec r)=\lambda{\delta(r-r_0)\delta(\theta-\theta_0)\over r}
=\lambda\delta^{(2)}(\vec r-\vec r_0)
\label{6}
\end{equation}
and

\begin{equation}
\vec J(\vec r)=j_0\hat z{\delta(r-r_0)\delta(\theta-\theta_0)\over r}
=j_0\hat z\delta^{(2)}(\vec r-\vec r_0),
\label{7}
\end{equation}
which correspond to the physical situations that we
have explained above.

Substituting (6) into (4), and (7) into (5),
we get the following linear self-energy densities

\begin{equation}
{U_{\rm Ele}\over \ell}={1/2}\lambda^2 G_p^{(2)}(\vec r_0,
\vec r_0)|_{Ren}
\label{8}
\end{equation}
and

\begin{equation}
{U_{Mag}\over \ell}={j_0^2\over 2c^2} G_p^{(2)}(\vec r_0,
\vec r_0)|_{Ren}.
\label{9}
\end{equation}

In order to obtain a finite result for the linear
self-energy densities given above, in the evaluation of
the Green's function we  adopt a renormalization
procedure to extract its irregular part, as indicated in our notation.
On the other hand, we have to be sure that no self-energy
survives in the absence of the topological defect.
So, we subtract from the Green's function its
Minkowski space-time correspondent and only then take the coincidence limit. In
this way, we have

\begin{equation}
G_p^2(\vec r_0,\vec r_0)|_{Ren} = \lim_{\vec r\to \vec r_0}
[G_p^{(2)}(\vec r,\vec r_0) - G_1^{(2)}(\vec r,\vec r_0)].
\label{10}
\end{equation}
As we shall see, this procedure gives us a well-defined expression
for equations (8) and (9). Now, all that we have to do,
is to obtain the conical geometry two-dimensional Green's function,
$G_p^{(2)}(\vec r,\vec r \,')$,
solution of the differential equation

\begin{equation}
\nabla_{(2)}^2 G_p^{(2)}(\vec r,\vec r \,')=-4\pi\delta^{(2)}
(\vec r-\vec r \,').
\label{11.a}
\end{equation}
$G_p^{(2)}(\vec r,\vec r \,')$, in the angular variables, presents periodicity
$2\pi\alpha$, whereas the two-dimensional Laplacian operator,
$\nabla_{(2)}^2$, in cylindrical coordinates reads

\begin{equation}
\nabla_{(2)}^2 ={1\over r}\partial_r r\partial_r +
{1\over r^2}{\partial^2_\theta}.
\label{11.b}
\end{equation}

For equation (11), we otained the two-dimensional Green's function

\begin{equation}
G_p^{(2)}(\vec r,\vec r \,') = -\ln
\left[r^{2p}+r'^{2p} - 2(rr')^p \cos p(\theta-\theta')\right],
\label{16}
\end{equation}
which, as we can see, presents in both angular variables a
periodicity $2\pi\alpha$ and reduces to the standard
two-dimensional Green's function for the $p=1$ case [10].

At this point, we are in position to obtain the electric
and magnetic linear self-energy densities by calculating
$G_p^{(2)}(\vec r_0,\vec r_0)|_{Ren}$ from (10):

\begin{equation}
G_p^{(2)}(\vec r_0,\vec r_0)|_{Ren} = -\lim_{\vec r\to\vec r_0}
\ln\left[{{r^{2p}+r_0^{2p} - 2(rr_0)^p\cos p(\theta-\theta_0)}\over
{r^2+r_0^2 - 2rr_0\cos (\theta-\theta_0)}}\right].
\label{17.a}
\end{equation}
This limit can be obtained, independently, by two
distinct ways:(i) taking $r=r_0$ and $\theta\to\theta_0$,
or (ii) $\theta=\theta_0$ and $r\to r_0$. In both cases
we get

\begin{equation}
G_p^{(2)}(\vec r_0,\vec r_0)|_{Ren} = -2[\ln p + (p-1)\ln r_0],
\label{17.b}
\end{equation}
which is well defined at the point $\vec r_0\ne 0$ and
vanishes for $p=1$; i.e., in the absence of the topological
defect.

For the linear self-energy densities we get

\begin{equation}
{U_{Ele}\over \ell} = -\lambda^2[\ln p + (p-1)\ln r_0]
\label{18.a}
\end{equation}
and

\begin{equation}
{U_{Mag}\over \ell} = -{j_0^2\over c^2}[\ln p + (p-1)\ln r_0] .
\label{18.b}
\end{equation}

The electrostatic self-force on the linear charge
distribution, induced by the lack of flatness of
this conical two-geometry, can be obtained by taking
the negative gradient of (16). Doing this, we get

\begin{equation}
{\vec F_{Ele}\over \ell} = -\vec \nabla {\Bigl({U_{Ele}\over \ell}\Bigr)}
= {{(p-1)\lambda^2}\over r_0} \hat r,
\label{19}
\end{equation}
which is a repulsive force for $p>1$, in agreement with
previous results [3,4] for a
charged point particle placed in this conical space-time.

The magnetostatic self-force cannot be obtained
from the linear self-energy in this way and, for this reason, we
adopt another procedure. (We shall discuss the magnetostatic
self-force in an energetic context at the end of
this paper.). The magnetostatic forces on an
arbitrary distribution of electric current densities,
can be obtained by the following expression:

\begin{equation}
\vec F_{Mag} = {1\over c}\int d^3r \vec J(\vec r)\times
\vec B(\vec r),
\label{20}
\end{equation}
where the magnetic field $\vec B(\vec r) = \vec \nabla
\times \vec A(\vec r)$. The vector potential, $\vec A(\vec r)$,  for a straight
wire carrying a constant current can be expressed, in the Coulomb gauge,by

\begin{equation}
\vec A(\vec r) = {1\over c}\int d^2r G_p^{(2)}(\vec r,
\vec r \,') \vec J(\vec r \,').
\label{21}
\end{equation}

Substituting (20) into (19), and after a few steps,
we get the  magnetic self-force per unit length

\begin{equation}
{\vec F_{Mag}\over \ell} = {1\over c^2} \int \int
d^2r d^2r'\bigl(\vec J(\vec r)\cdot \vec J(\vec r \,')\bigr)
\vec \nabla G_p^{(2)}(\vec r,\vec r \,').
\label{22}
\end{equation}
Finally, the expression for the linear
magnetic self-force density can be found by
substituting the current density,
$\vec J(\vec r)$
given by (7), into (21) yielding

\begin{equation}
{\vec F_{Mag}\over \ell} = {j_0^2\over c^2}
\vec \nabla G_p^{(2)}(\vec r_0,\vec r_0)|_{Ren}.
\label{23}
\end{equation}

Again, in order to obtain a finite well-defined
expression for (22) we have, as mentioned in our
notation, to adopt a renormalization procedure
similar to the previous one:

\begin{equation}
\vec \nabla G_p^{(2)}(\vec r_0,\vec r_0)|_{Ren} =
\lim_{\vec r\to \vec r_0} \vec \nabla
[G_p^{(2)}(\vec r,\vec r_0) -
G_1^{(2)}(\vec r,\vec r_0)].
\label{24}
\end{equation}

In calculating (28), we have to take first the
derivative  and then the coincidence limit.
Once again, the coincidence limit can be taken by
two different ways as explained before. In
both caes, we get for (28) the same result

\begin{equation}
\vec \nabla G_p^{(2)}(\vec r_0,\vec r_0)|_{Ren} =
-{(p-1)\over r_0}\hat r.
\label{25}
\end{equation}
Leading, consequently, to

\begin{equation}
{\vec F_{Mag}\over \ell} = -{j_0^2\over c^2}
{(p-1)\over r_0}\hat r,
\label{26}
\end{equation}
which is attractive for $p>1$.

A particular point which we decided to investigate
is how, from the previous expressions for the electric case, we can obtain the
magnetotastic
self-force  by
a Lorentz transformation.
In other words, we want to investigate what two
different inertial observers detect as the self-force of
a linear charge distribution, when one of them is
moving parallel to the distribution with velocity
$v=c\beta$ and the other one is at rest.
In order to answer this question we should take
an appropriate Lorentz transformation: a boost of the charge distribution along
the z-direction with coefficients $a_0^0 = -\gamma$
and $a_3^0 = -\beta\gamma$, where $\gamma =
{(1-\beta^2)}^{-1/2}$.
Taking this transformation, we get for the
moving observer, besides an electric charge distribution,
an electric current density. Hence, this observer
will detect an electric and a magnetic
self-force, respectively given by

\begin{equation}
{\vec F_{Elec}'\over \ell} = {{(p-1)\lambda^2}
\over {(1-\beta^2)r_0}}\hat r
\label{27.a}
\end{equation}
and

\begin{equation}
{\vec F_{Mag}'\over \ell} = -{{(p-1)\lambda^2\beta^2}
\over {(1-\beta^2)r_0}}\hat r.
\label{27.b}
\end{equation}
Therefore, the total (Lorentz) self-force
detected by this observer is

\begin{equation}
{\vec F_L'\over \ell} = {{(p-1)\lambda^2}
\over r_0}\hat r,
\label{27.c}
\end{equation}
in complete agreement
with the previous result given by Eq. (18), when the
observer is at rest in the original coordinate frame.

Before we finish this paper we would like to
make a few comments about our most important results for the conical
two-geometry with positive curvature; i.e., $p>1$:
as we have shown, the induced electrostatic self-force
on a linear charge density is repulsive, in contrast with the
magnetic self-force on the electric constant current,  which is
attractive. Although we have obtained
these results in an analytic way, we  could also
obtain a similar conclusion using the image method,
which is only applied for integer $p$. (See Ref.
[4].) We  can also understand these self-forces
by an energetic analysis: using Eqs.(16) and (17), we can see
that, although there may exist an arbitrary energy
background, which we did not consider,
the linear self-energy density decreases as
$r_0$ goes to infinity. For this reason, the
electrostatic self-force is repulsive. However,
for the magnetostatic case, the self-force cannot be
expressed as the negative gradient of the self-energy. In fact, we found that
the self-force is the positive gradient of the self-energy. This is not a
fortuitous coincidence: the explanation lies in the derivation of the
conservation of energy for a rigid circuit subjected to a virtual displacement
$d\vec r$, under the influence of the magnetic forces acting on it, keeping all
the currents constant. For this case, taking into account the work performed by
external sources against the induced electromotive forces, the change in the
magnetic energy is given by $dU=\vec F. d\vec r$ (this subject is exhaustively
discussed in  Ref. [11]).  Considering
these facts in  our analysis, and according to  Faraday's
law, it follows that the magnetostatic
self-force is expressed as
\begin{equation}
{\vec F_{Mag}\over \ell} =
{\vec \nabla\Bigl({U_{Mag}\over \ell}\Bigr)}.
\label{28}
\end{equation}
As we can see, the sign in front of the gradient
operator in Eq. (29) is positive, producing  the same
result obtained before in the explit calculation (see Eq.(25)).

With the calculations that we have done involving the self-force on a constant
current parallel to an infinite, straight cosmic string, we complete the
analysis of the self-forces on electromagnetic sources initiated by DeWitt and
DeWitt [12] in 1964. We believe that the results here presented can be of help
in understanding the behavior of a superconducting string [8] in the background
space-time of a cosmic string and the electromagnetic properties of disclinated
media [5,9].
\bigskip

\noindent
{\bf Acknowledgements}

This work was partially supported by CNPq and FINEP. We are indebted to the
referee for pointing out the references given in [10].
\bigskip

\noindent
{\large \bf References}

\bigskip

\noindent
\begin{enumerate}
\item A. Vilenkin, Phys. Rev. {\bf D23}, 852 (1981); W.A. Hiscock, Phys. Rev.
{\bf D31}, 3288 (1985); B. Linet, Gen. Rel. Grav. {\bf 17}, 1109 (1985).
\item S. Deser and R.  Jackiw, Comm. Math. Phys.
{\bf118}, 495 (1988);\\ Ph. Gebert and R. Jackiw, Comm. Math.
Phys. {\bf 124},  229 (1989);\\ V. B. Bezerra, Phys. Rev. {\bf
D34}, 3288 (1987).
\item B. Linet, Phys. Rev. {\bf D33}, 1833 (1986).
\item A. G. Smith, in {\em Proceedings of
Symposium on The Formation and Evolution of Cosmic
Strings}, edited by G. W. Gibbons, S. W. Hawking and
T. Vachaspati (Cambridge University Press, 1990)
\item Cl\'{a}udio Furtado and Fernando Moraes,
Phys. Lett. A {\bf 188}, 394 (1994).
\item M. O. Katanaev and I. V. Volovich, Ann.
Phys. (NY) {\bf 216}, 1 (1992).
\item The disclination can be obtained by
either removing (positive-curvature disclination),
or inserting (negative-curvature disclination)
a wedge in the solid.
\item E. Witten, Nucl. Phys. {\bf B249}, 557 (1989).
\item Cl\'audio Furtado et al., Phys. Lett. A {\bf 195}, 90 (1994).
\item The solution for the Green's function in a two-dimensional conical
space-time, has also been obtained by T. Souradeep and V. Sahni, Phys. Rev.
{\bf D46}, 1616 (1992) and M.E.X. Guimar\~aes and B. Linet, Class. Quantum
Grav. {\bf 10}, 1665 (1993); the apparent disagreament  by a constant factor of
$2\pi$ between their solution and ours is due to our use of CGS-Gaussian units
whereas they use MKS units. We also point out that a solution of the analogue
differential equation with another boundary condition was obtained a long time
ago by H. M. Macdonald, Proc. Lond. Math. Soc. {\bf 26}, 156 (1895).
\item W. K. H. Panovsky and M. Phillips, {\em Classical
Electricity and Magnetism} (Addison-Wesley Publishing Co.,
Massachusetts, 1962), 2nd ed. pg. 173.
\item C.M. DeWitt and B.S. DeWitt, Physics (N.Y.) {\bf 1}, 3 (1964).
\end{enumerate}

\end{document}